# Ionic Interactions in Biological and Physical Systems: a Variational Treatment

## Bob Eisenberg

*June 7, 2012*

Department of Molecular Biophysics and Physiology,
Rush University,
1750 West Harrison Street,
Chicago IL 60612, USA.
Fax: +312 942 8711;
Tel: +312 942 6467;
E-mail: beisenbe@rush.edu

# Abstract

Chemistry is about chemical reactions. Chemistry is about electrons changing their configurations as atoms and molecules react. Chemistry has for more than a century studied reactions as if they occurred in ideal conditions of infinitely dilute solutions. But most reactions occur in salt solutions that are not ideal. In those solutions everything (charged) interacts with everything else (charged) through the electric field, which is short and long range extending to the boundaries of the system. Mathematics has recently been developed to deal with interacting systems of this sort. The variational theory of complex fluids has spawned the theory of liquid crystals (or *vice versa*). In my view, ionic solutions should be viewed as complex fluids, particularly in the biological and engineering context. In both biology and electrochemistry ionic solutions are mixtures that are highly concentrated (~10 M) where they are most important, near electrodes, nucleic acids, proteins, active sites of enzymes, and ionic channels. $Ca^{2+}$ is always involved in biological solutions because the concentration (really free energy per mole) of $Ca^{2+}$ in a particular location is the signal that controls many biological functions. Such interacting systems are not simple fluids, and it is no wonder that analysis of interactions, such as the Hofmeister series, rooted in that tradition has not succeeded as one would hope. Here, we present a variational treatment of hard spheres in a frictional dielectric with the hope that such a treatment of an electrolyte as a complex fluid will be productive. The theory automatically extends to spatially nonuniform boundary conditions and the nonequilibrium systems and flows they produce. The theory is unavoidably self-consistent since differential equations are derived (not assumed) from models of (Helmholtz free) energy and dissipation of the electrolyte. The origin of the Hofmeister series is (in my view) an inverse problem that becomes well posed when enough data from disjoint experimental traditions are interpreted with a self-consistent theory.

# Introduction

Science flows in certain directions because scientists, like all people, are creatures of our training. Theories taught in our youth, before we learn critical skills, are hard for us to see objectively, and even harder to modify. But modify we must, as disciplines overlap, questions intermingle, and methods evolve. Modify we must if important questions remain unsolved despite the diligent work of generations of scientists.

**Chemistry**

Chemistry is first of all about chemical structures and reactions on the atomic scale. Chemists were excited by molecules and flows of electrons within those structures. Chemistry evolved without much thinking about macroscopic flow of molecules in space and time. Chemistry made remarkable advances because of this tight focus on chemical reactions. Physical chemists studied the tiny part of phase space that describes systems without macroscopic flow. The immense power of equilibrium thermodynamics and statistical mechanics[1-3] culminated in the theory of simple liquids, with its cathedrals[4-8] of analysis. The theory of simple liquids provides elegant insights of power and complexity. The founders of thermodynamics would surely have shared my view that the theory of simple liquids is an aesthetic and scientific masterpiece.

**Simulations**

In the last decades, aesthetics have often been replaced by calculations. Simulations have replaced much of the theory of simple liquids, judging by the number of papers published, but simulations have not yet dealt successfully with most macroscopic phenomena in my opinion[9], because in most cases simulations have not yet been able to confront macroscopic phenomena in the form found in the laboratory[10-47] and in chemical engineering every day.[48]

As the simulation profession adopts the standards of validation of the experimental community[48-50], it may well replace analysis as the best connector between atomic scale descriptions of chemical reactions and macroscopic measurements of the output of these reactions. But that has not happened yet, in my view and the consequences are serious.

Simulations are not mathematics, as much as I wish they were. Simulations cannot rely on the theorems that buttress and underlie numerical analysis. In my view, simulations need to be continually checked by comparison with test cases (e.g., conservation laws, model systems with known properties) just as do experiments. Until they fit experiments in conditions similar to those of interest, they have limited use. In the case of biological systems, those conditions are troublesome. All of life occurs in salt solutions derived from oceans. Without the proper ionic environment, cells burst, proteins coagulate, life stops. If simulations are not performed in solutions resembling those of life, they clearly cannot fit the properties of those solutions. Thus, simulations must be done in mixtures of $Na^+$, $K^+$, $Ca^{2+}$ and $Cl^-$ ions of substantial concentration (~0.5 M). Biological solutions are mixtures always involving $Ca^{2+}$ ions over a wide range of concentrations ($10^{-8}$ M as intracellular signals; ~$10^1$ M near and in active sites, nucleic acids, and ion channels). I am unaware of simulations done in these conditions, although surely some will be done soon enough. Until simulations reproduce the properties of life's solutions that are essential to maintain life, they are in a never-never land, in which their awesome detail and power can actually mislead. As any experimental biologist knows, uncalibrated measurements, or solutions of unknown composition can produce a bewildering array of results, that take more than a lifetime to sort out, most of which are artifacts, having little to do with biological function. One cannot expect uncalibrated simulations to do better than uncalibrated experiments.

**Mathematics**

Applied mathematics has a history quite different from chemistry. Applied math has always described macroscopic movement and flow. Indeed, mathematics of flow developed rapidly in the 1800's before molecules were thought to be real by many physicists.

Mathematics developed so powerfully that measurements of infinitesimal changes in fluids could be transformed into successful predictions of macroscopic flows. Infinitesimal measurements could be describe by infinitesimal (partial differential) equations (e.g., Navier Stokes) describing matter within boundaries. The meaning of the symbols in the equations could be forgotten, and the differential equations treated as abstractions. The abstractions could be integrated analytically and numerically (as mathematical objects without additional physical assumption).

An infinitesimal treatment could tell us what large masses of fluids do after the infinitesimals were integrated into macroscopic reality by mathematicians who knew nothing of what they were integrating.[51, 52] Water flowing in showers, water waves in oceans, even the action of wind on the water could be determined by mathematics. Indeed, computational fluid dynamics can predict complex flows (with shock waves, over complex structures) like those of air around a supersonic airplane, with a power and accuracy that leaves outsiders like me breathless.

**Mathematics is Different**

Mathematics learned to deal with certain classes of systems (conservative systems without dissipation) in a more abstract way, using the variational treatment of Hamilton. This variational approach had the important advantage of dealing with interactions automatically by algebra. The differential equations of motion were derived from the Least Action Principle. Components could be added to the Hamiltonian and the interactions of the components and the differential equations of motions re-derived for the composite system. The interactions of the new and old components appeared automatically in the differential equations without further physical argument. In this way, complex systems could be studied without ad hoc physical study of interactions, if the original and composite systems involved the same physics described by similar Hamiltonians.

The complex systems of the mathematicians did not involve chemicals, sad to say. The chemicals of the fluids and the atoms of the Hamiltonian were ignored by mathematicians, just as the interactions and macroscopic flow of the chemicals were ignored by the chemists, for many years. Chemicals were mysterious shadows to most mathematicians, as partial differential equations were mysterious moving shadows, a cinema of darkness, to most chemists.

**An interdisciplinary attempt to fill in the shadows**

I argue here that problems like the physics and physical chemistry of the Hofmeister series should be viewed in a new light, the additional light of mathematics, with a good measure of engineering insight thrown in. The light of just one discipline casts shadows too deep to see into. Light from another direction can fill those shadows substantially. Of course, the light of the other discipline creates its own shadows. So the trick is to use each discipline's light and ignore its shadows, melding this vision into a coherent self-consistent view of the entire phenomena. It should go without saying that I do not know how to do that, but I do think I know where to aim the light.

The essential difficulty in dealing with Hofmeister effects is, as I see it, that everything depends on interactions. To put it crudely, in the ideal noninteracting world, there is no Hofmeister series.

**Inverse Problems: reverse engineering the Hofmeister Series**

Of course, saying everything depends on interactions says too much, and says it too vaguely. There are many types of interactions and keeping them separate is essential if we are to avoid confusion. We need to keep the interactions if we wish to calculate what a solution would do from first principles.

The problem is much worse if we face it backwards, as I think we actually do. We are actually trying to infer the atomic basis of the Hofmeister series from macroscopic data taken under a wide range of conditions. This is an inverse problem, a problem in reverse engineering. Such problems are notoriously tricky because they are both ill-posed and over-determined, in the typical case. Sorting out the class of physical interactions is crucial if we are to make progress with our inverse problem.[53, 54] As difficult as the problems seem, there is real hope of progress, at least in the biological case, when the natural selection of evolution has created a reasonably robust device described by reduced equations,[55] as it apparently has when dealing with the selectivity of calcium[56, 57], sodium[58], and ryanodine receptor[59] channels.

Let's start with the simplest case. Let's talk about an implicit solvent model of ionic solutions, in which ions are hard spheres, of different charge and diameter, in a dielectric which also provides friction for flow. We consider the possibility that the Hofmeister effect might arise in even this oversimplified system[60] because of interactions that have not been dealt with self-consistently by previous methods. Of course, it is much more likely (from the chemists' perspective) that additional physics, beyond that of the implicit solvent primitive model will be needed to produce the effect. From the mathematicians' perspective, one cannot tell what a model is unless it is computed self-consistently, so the mathematician has no opinion of earlier work.

Consider a homogeneous system of say NaCl. We have spheres of two diameters and two charges. Monte Carlo simulations of this system do surprisingly well compared to real experimental data (see[18] for a recent treatment and references to the enormous literature). Monte Carlo simulations also do well for another monovalent salt say KCl. If we mix NaCl and KCl over a range of concentrations, difficulties arise because of interactions of the hard spheres within the physical model I just described. If divalents are involved—$CaCl_2$, for example—we start with pure simulations of pure $CaCl_2$, and then mix $CaCl_2$ with NaCl and/or KCl.

Mixed solutions have many interaction terms. Even if the physics in these mixtures is the same as in the pure solution (spatially uniform dielectric coefficient, etc.), the interactions are difficult to deal with. So many terms are involved, and theories and measurements have enough uncertainties, that it is very difficult to evaluate the coefficients of interactions.

In fact, mixed solutions of this type are very important to chemical engineers and have received a great deal of attention by them, summarized in[40, 43, 61]. Nonetheless, it seems clear that the many attempts to create a coherent self-consistent model of the activity (free energy per mole) of ionic mixtures has been a difficult one. A generation of physical chemists tried[38, 42, 43, 45-47, 62-69] culminating in the compendium of reference[43] that summarized the work of a taskforce of chemical engineers funded for many years for this purpose. Equations of state were awkwardly complex (summarized in[70], see also[71-78]). Apparently promising simplifications[40, 69, 79] seem not to have taken hold. Modern treatments and compendia of results for chemical engineers[29, 61] seem to lack physical basis or 'transferability'. That is to say, models and parameters developed for one type of mixture do not apply to other mixtures and other cases.

Enormous efforts of many workers (only some of which I know about[10-27, 30, 32-34, 59, 80-108]) have not produced a coherent view of ionic mixtures, containing divalents, as far as I can tell. This is not just a biologist's view, looking from the outside. A recent volume[29] described progress in dealing with specific properties of ionic solutions of many groups. An overview article by Kunz and Neuderer[28] concluded

> *"It is still a fact that over the last decades, it was easier to fly to the moon than to describe the free energy of even the simplest salt solutions beyond a concentration of 0.1M or so."*

This situation is rather upsetting, particularly to me as a biologist, mostly interested in biological solutions, because our solutions are those that Kunz and Neuderer say are not well described, let alone understood.

Biology occurs in salt solutions evolved from the oceans of the world. Biological salt solutions are invariably mixtures of NaCl and KCl with $CaCl_2$. $Ca^{2+}$ plays a particularly important role because the concentration of $Ca^{2+}$ acts as a specific control signal for many of life's processes. Indeed, $Ca^{2+}$ signals are as important and diverse in biology as electrical voltage is in computers! The quotation of Kunz and Neuderer implied what I feared[109-111]: the physical chemistry of life's solutions is not understood.

Here I suggest a different approach. I present a new variational treatment of ionic solutions[112-115] that produces (nearly) unique results for mixtures. I propose that this treatment can be used to systematically refine treatments of mixtures allowing a (hopefully) rapid convergence to a full understanding. But this is an approach not a result. The computations to compare with specific experiments have not been done. The feasibility of those computations has been shown, but the detail necessary to compare with results has not been computed.

**The variational approach**

The approach is to start with a simple model of a pure solution (say the primitive model of NaCl). Write it in a variational form. Derive the appropriate differential equations by the Euler Lagrange process. Learn to integrate those equations efficiently in a setup that allows evaluation of activities. Compare with experiment.

The approach repeats the same procedure with another pure solution, say $CaCl_2$ to be definite.

Then a variational model of a mixture of NaCl and $CaCl_2$ is made (by combining energy and dissipation) and converted into differential equations with boundary conditions appropriate for the experimental setup. The resulting differential equation has minimal adjustable parameters. Most coupling terms are determined by the Euler Lagrange process itself. The differential equation for the mixture is then integrated and applied in conditions that determine activity coefficients of the components of the mixture. Comparison is made with experiments.

When conditions are found in which the model fails to fit experiments, additional physics is included in the variational model. For example, a specific model of electron interaction between $Ca^{2+}$ and water is made, or a specific model of the spatial variation of dielectric coefficient is made, and included in the variational model. This model is then converted into a set of differential equations for mixtures just as was the primitive model, in the last few paragraphs.

**Ionic Solutions as Complex Fluids**

Ionic solutions are complex fluids because they contain charged particles. Ions in solution interact so strongly through the electric field that they always come 'in pairs' (within ~1 part in $10^{15}$). Ionic solutions are necessarily neutral or they explode.[116] Most ions also interact through their finite size.[23, 29, 35, 62, 63, 111] At biological concentrations, the shape of the ion dramatically distorts the shape of the electric field. The electric field of 1M spheres (approximately 8Å apart) of diameter 2Å is not well approximated by the electric field of 1M points. Finite size ions cannot occupy the same space, so ionic solutions are particularly complex in crowded confines near electrodes, active sites, and ion channels. Devices that use ions often depend on the nonideal properties of crowded ions. Ionic solutions are not simple fluids, particularly where they are most important, i.e. where they are crowded. Ionic solutions need to be treated as the complex fluids that they are, in my opinion.

It is best now to add some specifics to these abstractions. To do that, we consider a specific variational model of ionic solutions,[112-115, 117] that has grown from the theory of complex fluids[118-126] developed in the context of liquid crystals, [120, 122, 127-129] more than anything else.

The approach uses the dissipation principle as described by Liu[118] and co-workers built on the analysis of dissipative systems of Rayleigh [Strutt],[130-132] Onsager,[133-136] and Doi, explicitly[121, 137] and implicitly,[129] among many others[11, 119, 120, 122-126, 138-142]

We start with the key fact that the ions in water satisfy the dissipation law

$$\frac{d}{dt}\int \overbrace{\left\{k_B T \sum_{i=1}^{N} c_i \log c_i + \frac{1}{2}\left(\rho_0 + \sum_{i=1}^{N} z_i e c_i\right)\phi\right\}}^{\text{Helmholtz Free Energy Conservative Term}} d\vec{x}$$

$$= -\underbrace{\int \left\{\sum_{i=1}^{N} \frac{D_i c_i}{k_B T}\left|k_B T \frac{\nabla c_i}{c_i} + z_i q \nabla \phi\right|^2\right\} d\vec{x}}_{\text{Raleigh [Strutt] Dissipation Term}} \qquad (1)$$

This ensures that ions follow both an Energetic Variational Principle and also a set of simply derived differential equations. Either can be derived from the other.

$$\frac{dE}{dt} = -\Delta$$

or

$$\frac{\delta E}{\delta \vec{x}} = \tfrac{1}{2}\frac{\delta \Delta}{\delta \vec{u}};$$  (2)

or

$$\underbrace{\text{Conservative Force}}_{\substack{\text{from variation with respect to}\\\text{Position } \vec{x}}} = \underbrace{\text{Dissipative Force}}_{\substack{\text{from variation with respect to}\\\text{Velocity } \vec{u}}}$$

$E$ is the Helmholtz free energy; $\Delta$ is the Rayleigh[Strutt]/Onsager dissipation discussed at length in the literature.[112, 118, 121, 130-136, 138, 139]

I think it better physically to start with the variational principle because additional fields or components are best handled that way. The variational form of this expression allows easy derivation of the Euler Lagrange differential equations, which can also be written down directly in this case.

$$\frac{\partial c_i}{\partial t} = \nabla \cdot \left( D_i \left( \nabla c_i + \frac{z_i e}{k_B T} c_i \nabla \phi \right) \right),$$

(3)

$$\nabla \cdot (\varepsilon \nabla \phi) = -\left( \rho_0 + \sum_{i=1}^{N} z_i e c_i \right).$$

These are the usual Poisson Nernst Planck equations, called the Vlasov equations in plasma physics[143-145], drift diffusion in semiconductor physics[146-151] and PNP first[152] in biophysics[11, 55, 106, 111, 117, 153-212] and then in electrochemistry[158, 200, 213-215] and applied mathematics,[11, 106, 117, 146, 148, 149, 158, 159, 165, 186, 202-204, 206-209, 213, 214, 216] in which $c_i$ are the number densities of ions of species $i$, with diffusion coefficient $D_i$, valence $z_i$, charge $z_i e$ with $e$ the unit charge of a proton, diffusion (free) energy of $k_B T$, at electrical potential $\phi$, in permittivity $\varepsilon$, with protein permanent charge density $\rho_0$ as it represents the charge density of doping in semiconductors. $\rho_0$ depends on location and is zero in bulk solutions.

The coupling of these equations is tricky because the net charge $\sum_{i=1}^{N} z_i e c_i$ is well determined by the Laplacian of the electrical potential but cannot (in essence) be computed by summing the *individual* concentrations $c_i$ because the charge weighted sum is so tiny ($\sim 10^{-15}$) compared to individual concentrations. The system must be (nearly) electroneutral or potentials would strip electrons from atoms.

This treatment so far deals with point ions making a solution that is nonideal in a simple way. Interactions are only electrostatic, as in classical Debye-Hückel and Poisson Boltzmann equilibrium theories. This is not the primitive model.

**Spatially Nonuniform Boundary Conditions**

The variational method and the resulting Poisson Nernst Planck (PNP)11, 55, 106, 109, 117, 146-148, 151-153, 155, 159, 163, 164, 169, 170, 172-175, 179, 183, 184, 188, 190, 193, 194, 197, 201, 202, 204, 206, 207, 209-212, 214-219 equations have substantial advantages, however, over the traditional equilibrium approach. The PNP equations allow spatially nonuniform boundary conditions such as those needed to drive flux. In our electronic technology those spatially nonuniform boundary conditions are the power supply potentials. In biology, those spatially nonuniform boundary conditions are the electrical potential and concentration gradients across the cell membrane. In chemistry, those spatially nonuniform boundary conditions are the concentrations of the classical Nernst equation of electrochemistry and polarography, and perhaps an applied electrical potential as well.

A great deal is gained by using PNP as a nonideal theory allowing flow: all of the components of a computer can be built using circuit elements that satisfy PNP.148-151, 220-231 Resistors, capacitors, and transistor switches, transistor amplifiers, etc. are actually mathematical solutions of the PNP equations. Thus, moving to PNP allows one to write "all human knowledge and logic" (that can be performed by a computer) as a solution of a partial differential equation in a spatially complex domain with spatially nonuniform (but otherwise simple) boundary conditions.

**Finite Size of Ions**

To deal with the finite size of ions, and extend PNP into a theory of ionic solutions, we need to decide how to describe the finite diameter of the ions. This decision is surely not unique,[10-12, 15, 16, 19, 22-25, 27, 100, 103, 104, 106, 232] and the choice of description depends on the feasibility of calculations and the accuracy of fit of the resulting model to a broad range of data, from different types of experiments, in a broad range of solutions and concentrations. This work has not yet been done. The purpose of this paper is to facilitate and motivate such work.

We have included finite size at the length scale of PNP by adding terms to the above equations. The resulting model is variational, so it is a model in the spirit of the theory of complex fluids, but the fluid is the primitive model of electrolytes, leading to peculiar oxymoronic names for our model, like 'complex primitive theory' of electrolytes! Given the present state of checking of our models, (oxy)moronic names are perhaps appropriate.

To be specific, our variational theory describes the (Helmholtz free) energy and dissipation of ionic solutions by

$$\frac{d}{dt}\int\left\{k_B T\sum_{i=n,p}c_i\log c_i+\tfrac{1}{2}\left(\rho_0+\sum_{i=n,p}z_i e c_i\right)\phi+\sum_{i,j=n,p}\frac{c_i}{2}\int\tilde{\Psi}_{i,j}c_j d\vec{y}\right\}d\vec{x}$$

$$\text{Conservative}$$

(4)

$$=-\int\left\{\sum_{i=n,p}\frac{D_i c_i}{k_B T}\left|k_B T\frac{\nabla c_i}{c_i}+z_i e\nabla\phi-\sum_{j=n,p}\nabla\int\tilde{\Psi}_{i,j}c_j d\vec{y}\right|^2\right\}d\vec{x}$$

$$\text{Dissipative}$$

$\widetilde{\Psi}_{i,j}$ represents the Lennard Jones crowded charges terms defined in references[112, 118]. The variational formulation and variational derivatives are computed and discussed in those papers.

If ions are modeled as Lennard Jones spheres, the variational principle produces 'Euler Lagrange' equations of a drift-diffusion theory with finite sized solutes that are a generalization and correction of PNP.

$$\frac{\partial c_n}{\partial t} = \nabla \cdot \left[ D_n \left\{ \nabla c_n + \frac{c_n}{k_B T} \left( z_n e \nabla \phi - \int \frac{12\varepsilon_{n,n}(a_n + a_n)^{12}(\vec{x}-\vec{y})}{|\vec{x}-\vec{y}|^{14}} c_n(\vec{y}) d\vec{y} \right. \right. \right. \\ \left. \left. \left. - \int \frac{6\varepsilon_{n,p}(a_n + a_p)^{12}(\vec{x}-\vec{y})}{|\vec{x}-\vec{y}|^{14}} c_p(\vec{y}) d\vec{y} \right) \right\} \right] \quad (5)$$

always combined with the Poisson Equation

$$\nabla \cdot (\varepsilon \nabla \phi) = -\left( \rho_0 + \sum_{i=1}^{N} z_i e c_i \right) \qquad i = n \text{ or } p \quad (6)$$

We write the equation only for negative monovalent ions with valence $z_n = -1$ to keep the formulas reasonably compact. Programs have been written for all valences. $c_{p,n}(\vec{y})$ is the number density of positive $p$ or negative $n$ ions at location $\vec{y}$. $\varepsilon_{n,n}$ and $\varepsilon_{n,p}$ are coupling coefficients. $a_{p \text{ or } n}$ are the radii of ions.

**Field Theory of Ionic Solutions.**

These equations form a field theory of electrolytic solutions and so pretend to explain all phenomena of those solutions, at equilibrium and in flow. They automatically are self-consistent and change form as new ions are added into the system. Thus, these equations form a necessarily self-consistent model of the flow of mixtures of ionic solutions. Up to now, most such models have been rather *ad hoc*, and one could not be sure if they were self-consistent or not.[44, 233-253] Additional fields[254] (e.g., heat flow[138] or convection[112, 115]) can be added to our variational model by including their (Helmholtz free) energy in the energy functional $E$ and the dissipation in the dissipation functional $\Delta$ of equation (2).

Equations of this generality can be misleading, however, because difficulties often appear in the numerical implementation. Work to date shows that the calculations needed to describe electrochemical cells and ion channel are feasible and not particularly irksome, but it took some years to reach that conclusion. Each experimental setup requires translation into an approximate model, and then numerical procedures must be designed that focus work on the important features of the model these difficulties can be overcome one by one, as they have been in other models of the experimental setups.

**Applications to Specific Problems**

The field theory of ionic solutions has been applied to a number of problems of biochemical and chemical interest. The original publications[112, 113] computed curves of binding selectivity in two classical channels of considerable biological interest the calcium channel of cardiac muscle[57, 255-268] and the voltage activated sodium channel of nerve.[269-277]

Both were represented by the implicit solvent model with side chains represented as spheres free to move within the channel selectivity filter, but not able to move out of that region.[56-58, 111, 267, 278, 279] This model has proven remarkably successful in dealing with the important selectivity phenomena of these channels. A single model, with one set of parameters that are never changed, using crystal radii of ions, and one dielectric coefficient and one channel diameter, is able to account for selectivity data in a wide range of solutions (over 4 orders of magnitude of calcium concentration, and in solutions of varying $K^+$, $Na^+$, $Rb^+$ and $Cs^+$ concentration, for example). The calcium channel is represented by side chains Glu Glu Glu Glu and the sodium channel by Asp Glu Lys Ala. This work is reported in some 35 papers using various numerical methods including the variational model described here. It has been possible to invert the procedure. The inverse problem of determining the distribution of side chains inside the channel from current voltage relations in a range of solutions has actually been explicitly solved, using established methods of inverse problems, including the effects of noise and systematic error.[55]

The variational method allows calculations of current vs. time and current vs. voltage not previously performed on this reduced model of calcium and sodium channels. Such calculations are of some significance since they confront the measurements performed by hundreds of membrane biologists following the paradigms introduced by Hodgkin and Huxley, reviewed in reference[280].

The variational method was extended[115] to allow computation of the combined effects of water flow and electrodiffusion that underlie the function of many tissues and cells, from kidney, to epithelia, to cerebrospinal fluid and so. Indeed, classical physiology from Harvey to Pappenheimer[281, 282] might be considered the study of the flow of water and ions. As far as we know, this paper represents the first selfconsistent treatment of this classical subject, allowing cell volume changes, convection, and ionic electrodiffusion.

The application of the variational method to classical problems of physical chemistry has started with the classical charged wall problem. There we have shown[283] direct comparisons with the Metropolis Monte Carlo calculations of Henderson and Boda[86, 92, 284-289], which form a significant fraction of our understanding of ionic solutions in homogeneous and inhomogeneous situations.

The most striking success of these reduced models has been the description of the main calcium channel that controls contraction in muscle (and appears in neurons and many other cell types as well, so far with unknown function), the Ryanodine Recptor Ryr.[290-297] Gillespie has shown that an extension of the reduced model called PNP-DFT[185, 298, 299] does remarkably well[59, 268, 300-303] in predicting experiments of some complexity and subtlety (i.e., anomalous mole fraction effects and three cation mixtures) as well as drastic mutations changing charge densities from some 13 molar to zero. The theory used was developed[185, 297, 298] before selfconsistent variational methods were known to me, and is not selfconsistent. It leaves out the effects of changes in shape of the ionic atmosphere thought to be important 'forever',[44, 235-240, 247, 249, 304] and now found in textbooks.[35, 36, 305] These effects may in fact be unimportant in applications to channels, where the shape of the ionic atmosphere may be mostly determined by a (rather) unchanging protein. But the lack of self-consistency in PNP-DFT prevents it from being a good model of ionic solutions in physical chemistry, where changes in shape of the ionic atmosphere are thought to be dominant.[44] The variational treatment of ionic conductance[112, 113, 115, 118, 283] makes no simplifying assumptions about the shape of the ionic atmosphere: everything (that is in the model) interacts with everything else, to minimize the dissipation and (Helmholtz free energy) of the model.

This work is more practical science than vague theory. It is meant to connect the mathematical results of the theory of complex fluids with the practical problems of experimental biophysicists and physical chemists. Sadly, the authors know much less of those problems in physical chemistry than in biophysics, in which they have worked for many years. Thus, the utility of this approach for classical physical chemistry remains to be seen. Applications to the Hoffmeister series, and the experiments that define the Hoffmeister series are one important part of our approach to classical physical chemistry.

**Experimental Setups**

The experimental setups used to measure the Hofmeister series are varied and in my view must be understood in some detail before they can be well modelled by this new approach. The subtleties are nontrivial, as can be seen in the magnificent treatise of Hünenberger and Reif[16] which describes thousands of papers using hundreds of methods (I estimate) to determine the free energy per mole of individual ions. Many different setups are used and each requires a separate idealization, model, and mathematical description. In my view, theories should be tested against idealizations of experimental setups and not against metaphorical abstractions like thermodynamic 'theorems'. Too often, the theorems of thermodynamics and statistical mechanics have hidden assumptions in my personal opinion. Fitting data to metaphors is likely to cause even more confusion in that case, compounding the felony, as the legal saying goes. Fitting to real data avoids this problem, at the cost of some complexity in analysis.

Application of our variational model to the Hofmeister series requires models and calculations of the free energy per mole of individual ions. Those calculations will need the help of workers familiar with the experimental setups. As the primitive model for the Hofmeister series[60]—presented here in variational clothing—fails to fit some data, as it surely will, additional physics will no doubt need to be added, and the less primitive models will then have to be compared to experiments. It is not at all clear how much improvement a fully self-consistent model will provide at equilibrium in solutions of one salt. One may expect more improvement when interactions are larger and more complex, in concentrated environments, in mixtures, and in divalents. Of course, those are just the characteristics of biological solutions, so one can expect variational models to be most important as they are applied to living systems.

Along the way, the variational method will also allow introduction of boundary conditions and nonequilibrium properties that arise from spatially nonuniform concentrations and voltages. In my view, this is an important social and scientific advantage. Socially, this unites the classical thermodynamic approach (which had been isolated in its thermodynamic limit) with the classical field theories of the rest of physics. Scientifically, the nonequilibrium systems have properties of enormous importance that provide, for example, our modern electronic technology, and also allow voltage signals ('action potentials') to propagate along our nerve fibres.

**Specific Plan to implement a variational approach**

I present a specific plan that might implement these ideas. I am very aware that it is dangerous and presumptuous to present any plan for science. Science proceeds best when it is initiated and energized by the curiosity of individuals. The following is meant to provoke discussion to test the utility of the variational approach.

1) identify a specific set of experimental measurements that operationally define 'the Hofmeister Effect'
2) make a physical model of the setup of those experiment.
3) make a model of how estimates of parameters are made from the measurements of that experiment
4) make a mathematical model of the physical model, in the spirit of the variational theory of complex fluids
5) derive the field theory (Euler Lagrange) equations of the model
6) solve the equations of the model for the experimental boundary conditions.
7) compute estimates of parameters from the mathematical solutions of the equations, treating the output of the equations as if they were outputs from the experimental setup.
8) compare with experiments in a range of solutions and concentrations.
9) revise the model to include more and different physics to see if the fit to experiments is improved.

### Chemical Conclusions

Chemistry has focused on atoms and molecules with less concern about the surrounding environment, understandably enough, since mathematics was not available to deal with systems in which 'everything' interacts with 'everything' else. The solutions containing these atoms and molecules were treated as ideal fluids, in which nothing interacts with nothing, if my lapse in grammar may be excused. The variational theory of complex fluids has successfully dealt with systems in which everything interacts with everything else over many scales.

### Chemical Reactions in the Liquid State

In my admittedly biased view, chemical reactions in the liquid state should be analysed in the framework of complex fluids. Chemical reactions are interactions of reactants (microelements in the language of the theory of complex fluids) involving rearrangements of the internal (electronic) structure of reactants, according to the wave equation of the electron, Schrödinger's equation. Daunting interactions of electrons, microelements, and the macroscopic world can be handled automatically and self-consistently in the theory of complex fluids that cannot easily be handled by any other method I know of.

### Future History of Physical Chemistry

The future history of physical chemistry will deal with ionic solutions as complex fluids, I believe. When mixtures and divalents and flow are involved, the variational approach to complex fluids is likely to lead to substantial progress. The crucial step is to force the theory to confront the actual properties of such solutions as measured with many techniques under many conditions. Determining an appropriate model of ionic solutions is an inverse problem that requires masses of disjoint data, and a variational theory, to provide useful well-posed results.

# Acknowledgement

This paper is possible because of the support of others. Chun Liu, more than anyone else, has developed the energetic variational approach to the theory of complex fluids (and liquid crystals) and spared no effort to help me understand how we can apply it to ions in liquids and channels. The physical chemistry community in general, represented in the present instance by Pavel Jungwirth, has welcomed my view[9, 109-111, 266, 306] of long standing approaches in a most generous way, in the best of the academic tradition, with criticism and scepticism to be sure, but always listening. The computational electronics community[195, 307-317] has been similarly generous. Biophysicists have not always been like that.[9, 109-111, 163, 164, 171, 177, 266, 306, 318-321] Support was from Rush Medical College and I am particularly grateful for the administrative tasks the Dean's Office has not asked me to do.